\documentclass{amsart}
\usepackage{amsmath,amssymb,amsxtra,amsthm,amscd}
\usepackage[mathscr]{eucal}
\usepackage{bbm}
\usepackage[all,cmtip]{xy}
\usepackage{lmodern}
\usepackage{microtype}
\usepackage{graphicx}
\usepackage{todonotes}

\usepackage{clrscode}
\usepackage{multirow}

\newif\ifShowLabels
\ShowLabelstrue
\newcommand{\TeXref}[1]{
\marginpar{\scriptsize \texttt{#1}}}
\ShowLabelsfalse

\newcommand{\Parfor}{\kw{parallel for}}

\DeclareMathOperator{\C}{\mathbf{C}}

\DeclareMathOperator{\coker}{coker}

\DeclareMathOperator{\F}{\mathbf{F}}

\DeclareMathOperator{\Hom}{Hom}

\DeclareMathOperator{\M}{\mathbf{M}}

\DeclareMathOperator{\Os}{\mathbf{O-single}}
\DeclareMathOperator{\Od}{\mathbf{O-double}}

\DeclareMathOperator{\Sh}{Sh}

\DeclareMathOperator*{\one}{1}
\newcommand{\onehatplace}[1]
{ \one^{\substack{#1 \\ \frown}} }

\DeclareMathOperator*{\bones}{\times}
\newcommand{\undertimes}[1]
{ \bones_{#1} }

\DeclareMathOperator*{\bowl}{\cup}
\newcommand{\undercup}[1]
{ \bowl_{#1} }

\DeclareMathOperator*{\arch}{\cap}
\newcommand{\undercap}[1]
{ \arch_{#1} }

\newcommand{\pull}
{\!\!\! -\!\!\! -\!\!\! -\!\!\!}

\DeclareMathOperator*{\holimprep}{holim}                       
\newcommand{\holim}[1]%
{\displaystyle\holimprep_{\substack{\leftarrow \pull - \\ #1}} \, }

\DeclareMathOperator*{\hocolimprep}{hocolim}                   
\newcommand{\hocolim}[1]%
{\displaystyle\hocolimprep_{\substack{- \pull \rightarrow \\ #1}} \, }

\DeclareMathOperator*{\plainlim}{lim}                           
\newcommand{\contralim}[1]%
{\displaystyle\plainlim_{\substack{\leftarrow \pull - \\ #1}} \, }

\DeclareMathOperator*{\plaincolim}{colim}                       
\newcommand{\colim}[1]%
{\displaystyle\plaincolim_{\substack{- \pull \rightarrow \\ #1}} \, }

\DeclareMathOperator*{\laxlimplain}{laxlim}                     
\newcommand{\laxlim}[1]%
{\displaystyle\laxlimplain_{\substack{\leftarrow \pull - \\ #1}} \, }

\providecommand{\bysame}{\makebox[3em]{\hrulefill}\thinspace}

\theoremstyle{plain}
\newtheorem{Thm}{Theorem}[section]

\newtheorem*{MainThm}{Main Theorem}

\newtheorem{Cor}[Thm]{Corollary}

\newtheorem{Lem}[Thm]{Lemma}
\newtheorem{Prop}[Thm]{Proposition}

\theoremstyle{definition}
\newtheorem{Def}[Thm]{Definition}
\newtheorem{Alg}[Thm]{Algorithm}

\newtheorem{Ex}[Thm]{Example}

\newtheorem{Rem}[Thm]{Remark}

\theoremstyle{remark}
\newtheorem{Not}[Thm]{Notation}

\newtheoremstyle{freestylethm}{6pt}{6pt}{\itshape}{}%
                {\bfseries}{}{.5em}{\thmnote{#3}}
\theoremstyle{freestylethm}

\newcommand{\SecRef}[2]{\section{#1}\label{S:#2}%
\ifShowLabels \TeXref{{S:#2}} \fi}

\newcommand{\refS}[1]{\textup{\ref{S:#1}}}

\newcommand{\refT}[1]{\textup{\ref{T:#1}}}

\newcommand{\refD}[1]{\textup{\ref{D:#1}}}
\newcommand{\refC}[1]{\textup{\ref{C:#1}}}

\newcommand{\refR}[1]{\textup{\ref{R:#1}}}
\newcommand{\refN}[1]{\textup{\ref{N:#1}}}

\newenvironment{ThmRef}[1]%
{ \begin{Thm} \label{T:#1}
\ifShowLabels \TeXref{T:#1} \fi }%
{ \end{Thm} }
\newenvironment{DefRef}[1]%
{ \begin{Def} \label{D:#1}
\ifShowLabels \TeXref{D:#1} \fi }%
{ \end{Def} }
\newenvironment{LemRef}[1]%
{ \begin{Lem} \label{L:#1}
\ifShowLabels \TeXref{L:#1} \fi }%
{ \end{Lem} }
\newenvironment{CorRef}[1]%
{ \begin{Cor} \label{C:#1}
\ifShowLabels \TeXref{C:#1} \fi }%
{ \end{Cor} }
\newenvironment{RemRef}[1]%
{ \begin{Rem} \label{R:#1}
\ifShowLabels \TeXref{R:#1} \fi }%
{ \end{Rem} }
{ \begin{Prop} \label{P:#1}
\ifShowLabels \TeXref{P:#1} \fi }%
{ \end{Prop} }
\newenvironment{ExRef}[1]%
{ \begin{Ex} \label{E:#1}
\ifShowLabels \TeXref{E:#1} \fi  }%
{ \end{Ex} }
\newenvironment{NotRef}[1]%
{ \begin{Not} \label{N:#1}
\ifShowLabels \TeXref{N:#1} \fi }%
{ \end{Not} }

{ \begin{Thm} [#2]\label{T:#1}
\ifShowLabels \TeXref{T:#1} \fi }%
{ \end{Thm} }
{ \begin{Def} [#2]\label{D:#1}
\ifShowLabels \TeXref{D:#1} \fi }%
{ \end{Def} }
{ \begin{Lem} [#2]\label{L:#1}
\ifShowLabels \TeXref{L:#1} \fi }%
{ \end{Lem} }
{ \begin{Cor} [#2]\label{C:#1}
\ifShowLabels \TeXref{C:#1} \fi }
{ \end{Cor} }
\newenvironment{RemRefName}[2]%
{ \begin{Rem} [#2]\label{R:#1}
\ifShowLabels \TeXref{R:#1} \fi }%
{ \end{Rem} }
{ \begin{Prop} [#2]\label{P:#1}
\ifShowLabels \TeXref{P:#1} \fi }%
{ \end{Prop} }
{ \begin{Ex} [#2]\label{E:#1}
\ifShowLabels \TeXref{E:#1} \fi }%
{ \end{Ex} }

\begin{document}

\title[Singular Persistent Homology]{Singular Persistent Homology with Geometrically Parallelizable Computation}
\author{Boris Goldfarb}
\address{Department of Mathematics and Statistics,
State University of New York, Albany, NY 12222, USA}
\email{bgoldfarb@albany.edu}



\date{\today}

\begin{abstract}
Persistent homology is a popular tool in Topological Data Analysis. 
It provides numerical characteristics of data sets which reflect global geometric properties. In order to be useful in practice, for example for feature generation in machine learning, it needs to be effectively computable.   Classical homology is a computable topological invariant because of the Mayer-Vietoris exact and spectral sequences associated to coverings of a space. 
We state and prove versions of the Mayer-Vietoris theorem for persistent homology under mild and commonplace assumptions. 
This is done through the use of a new theory, the singular persistent homology, better suited for handling coverings of data sets.   
As an application, we create a distributed computational workflow where the advantage is not only or even primarily in speed improvement but in sheer feasibility for large data sets. 
\end{abstract}

\maketitle

\tableofcontents

\SecRef{Motivation and Statement of Results}{MT}

In order to describe the problem addressed in this paper, we start by recalling the standard treatment of persistent homology.  This construction is designed to leverage computations in algebraic topology in order to quantify geometric properties of finite data sets. 
It provides a multi-scale representation of geometric features of the data, including the relations between the scales.
On a more sophisticated level, one can use filtrations of a simplicial complex in place of scales to build a similar persistent homology signature of a filtration. 
Persistent homology has established itself as a robust feature included in machine learning applications in addition to other ad hoc uses in data science.
We refer to the surveys Edelsbrunner/Harer \cite{EH} and Carlsson \cite{gC:14} for expositions.

Consider a filtration of a general topological space $X$ with the indexing set $[0, +\infty)$.  This filtration can be thought of as a covariant functor $\phi$ from $[0, +\infty)$ with the usual total ordering to the category of subspaces of $X$ partially ordered by inclusion.  Usually in the applications the topological space is a simplicial complex, and the filtration stages are simplicial subcomplexes. Let us use the notation $X_{\varepsilon}$ for $\phi (\varepsilon)$. For a functor $H_n$ from the appropriate category to modules over a ring $R$, one has the induced diagram with the nodes $H_n (X_{\varepsilon})$ and the maps $H_n (X_{\varepsilon} \to X_{\varepsilon'})$ for any pair of non-negative reals $\varepsilon \le \varepsilon'$.  This whole diagram is sometimes referred to as the \textit{$H_n$-persistence module} of the filtration and can be denoted as $pH_n (\phi)$. 

The \textit{$\delta$-persistent $H_n$-module of $X_{\varepsilon}$} is the image of the induced map $H_n (X_{\varepsilon}) \to H_n(X_{\varepsilon + \delta})$. 
The \textit{$\delta$-persistent $n$-th Betti number} of $X_{\varepsilon}$ can be defined as the rank of this submodule.
This framework is then used to characterize the $H_n$-based features from 
$X_{\varepsilon}$ that are still present in $X_{\varepsilon + \delta}$.  For example, one can measure the \textit{survival interval} for each of the elements in $H_n (X_{\varepsilon})$ as the difference between the infimum of all $\varepsilon'$ such that the element is not in the kernel of the map $H_n (X_{\varepsilon}) \to H_n(X_{\varepsilon'})$ and the value of $\varepsilon$ itself.  

In order to describe the appearance of persistence modules in data science, we specialize to a finite metric space $M$ with $t$ points.  This is the usual framework for measuring dissimilarity between $t$ data points.  Associated to this metric space is the simplex $X$ spanned by all $t$ points as vertices.  

\begin{DefRef}{RIPS}
	The \textit{Vietoris-Rips filtration} $\rho$ of $X$ is by subcomplexes $X_{\varepsilon}$ defined by the condition that a subset $S$ of $M$ spans a simplex in $X_{\varepsilon}$ if and only if $d(p, q) \le \varepsilon$ for every pair of points from $S$. 
\end{DefRef}

Using the simplicial $n$-dimensional homology functor $H_n$ gives a persistent module for each dimension $n$.  In this case 
the $\delta$-persistent homology module of $X_{\varepsilon}$ characterizes the $n$-cycles in $X_{\varepsilon}$ that are not the boundary of any $(n+1)$-chain from the larger subcomplex $X_{\varepsilon + \delta}$.  In order to make these invariants computable in practice and explainable to a broad population of data scientists, it is prudent to specialize to fields as coefficients.  It becomes possible to use phrases such as ``the $\delta$-persistent Betti number of $X_{\varepsilon}$ measures the number of $n$-dimensional holes in $X_{\varepsilon + \delta}$ created by the subcomplex $X_{\varepsilon}$.''

\begin{NotRef}{NAYVD}
	When $M$ is a finite metric space, and the Vietoris-Rips filtration is understood, we will use the notation $H_n^{\varepsilon} (M)$ for $H_n (X_{\varepsilon})$ and the notation $pH_n (M)$ for the resulting persistence module $pH_n (\rho)$.
\end{NotRef}

Continuing with the setting in the last paragraph, there is a nice visual summary that can be constructed to describe the persistent module.  This description requires one to use structural theorems for finitely generated graded $K[x]$-modules, where $K$ is a field, so we will assume that the coefficient ring is a field in the rest of this paper.

The \textit{persistence barcode} of $M$ is a collection of intervals indicating the birth of a new generator in some $H_n^{\varepsilon} (M)$ and its death in $H_n^{\varepsilon + \delta} (M)$, where $\delta$ is the survival interval for that generator.  We refer to \cite{gC:14} for details and will include some illustrations shortly in this section.

Simplicial homology is one of the most computable algebraic invariants in topology.  There are two reasons for that.  One is that homology depends on a discrete representation of the space, its triangulation.  This allows to use linear algebra to perform computations over fields  whose complexity grows with the size of the triangulation but which use well understood and refined algorithms.  The other reason is the excision property.  This property is crucial for the axiomatic characterization of homology and distinguishes homology from other topological invariants.  

This goal of this paper is to state and verify an analogue of the Mayer-Vietoris property for persistent homology.  The goal of such a theorem should be to recover information about the persistent homology barcodes for the total data set from the barcodes associated to subsets that form a specific kind of covering.  Of course, in practice this kind of approach to computation is very desirable because it allows to parallelize the local computations and therefore makes persistent homology computations feasible for larger data sets under the same memory  constraints.  It is also possible that this divide-and-concur strategy would be more efficient than direct homology computations with $M$. 

We will use the rest of this section to demonstrate that this goal is not achievable in its entirety and offer a compromise applicable and sufficient in many practical situations.  
The fact that there is no perfect excision property for persistent homology shouldn't be surprising.  Analogues of purely topological theorems usually fail without careful incorporation of local information or preservation of relevant information about the intersections as in excision results for schemes in algebraic geometry.

\begin{figure}[h!]
      \centering
       \includegraphics[width=0.35\linewidth]{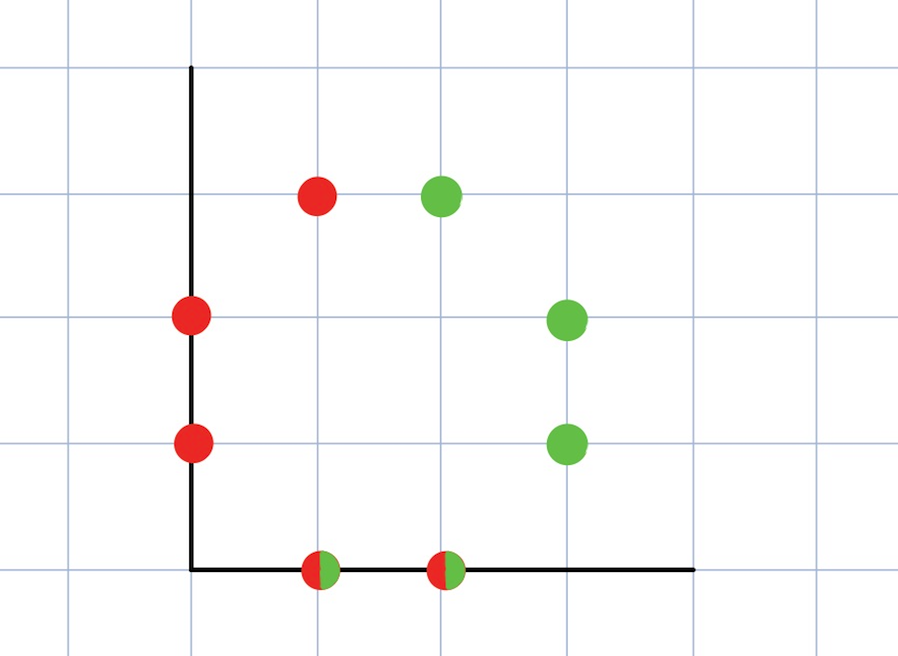} 
       \includegraphics[width=0.95\linewidth]{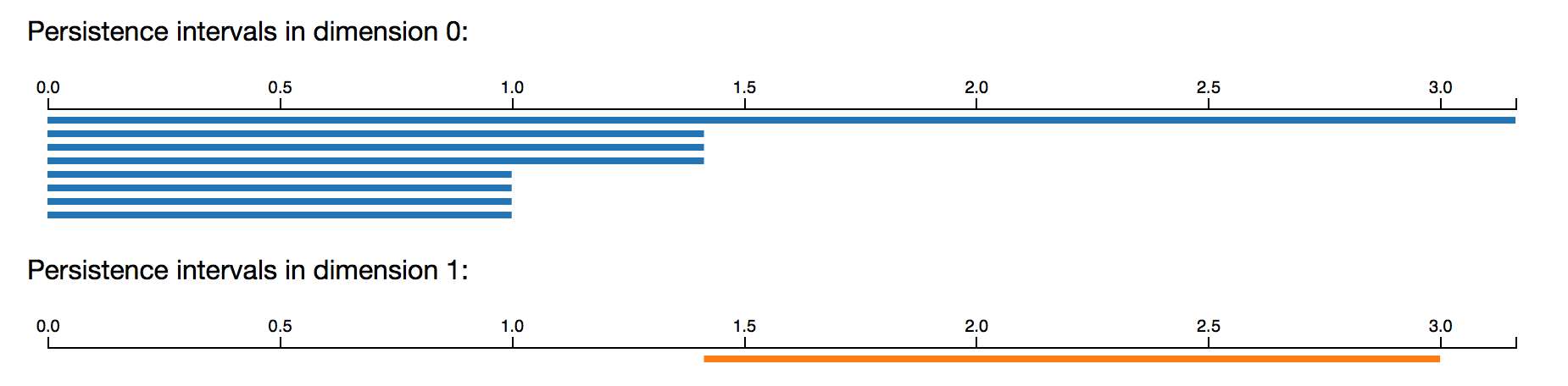}
       \includegraphics[width=0.55\linewidth]{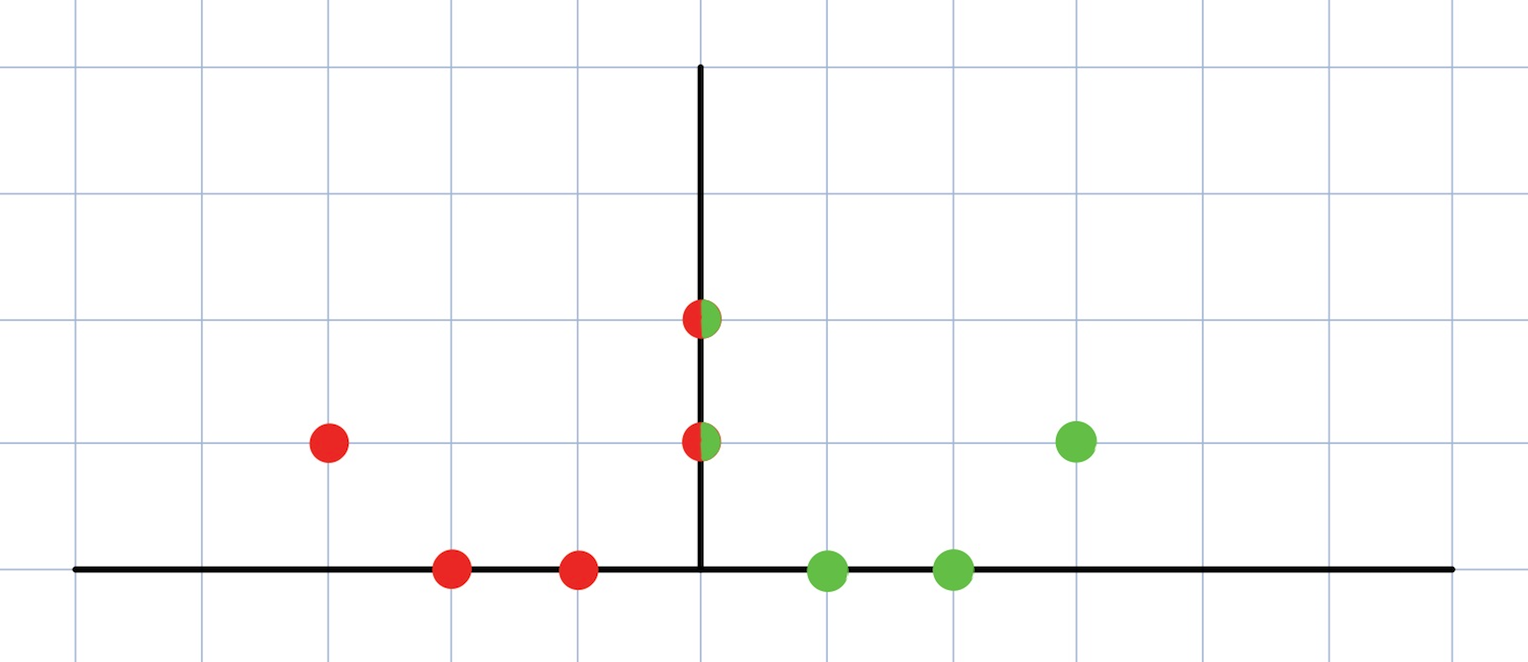}
       \includegraphics[width=0.95\linewidth]{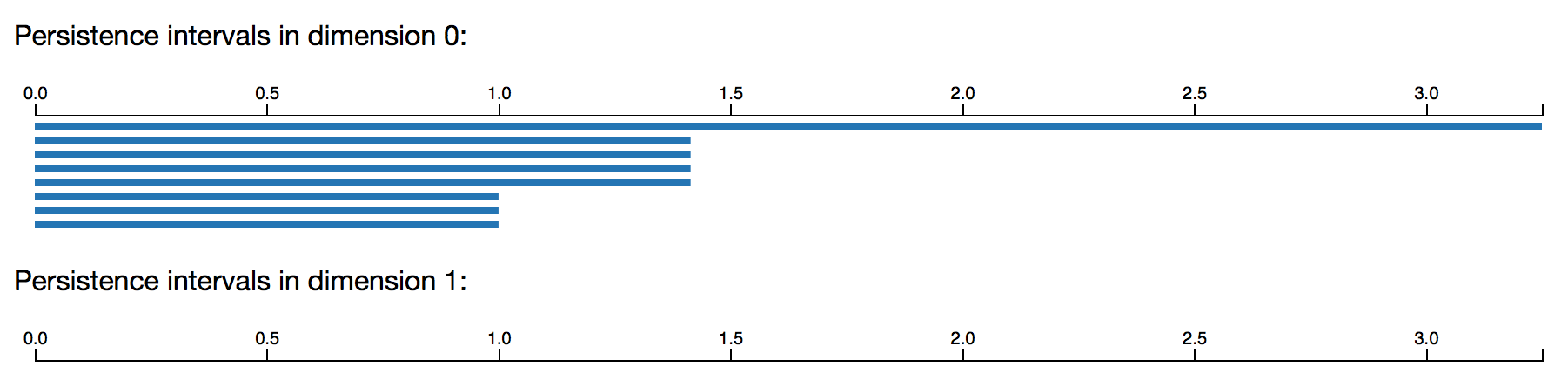}
       \caption{Reconstruction of the persistence barcode from a complete metric covering is not a well-posed problem.}
      \label{Counterexample}
   \end{figure}

\begin{ExRef}{MORE}
Here we compare persistence barcodes for two data sets in $\mathbb{Z}^2$: 
\begin{align*}
A &= \{ (0,1),
(0,2),
(1,0),
(1,3),
(2,0),
(2,3),
(3,1),
(3,2) \}, \\ 
B &= \{ (-3,1),
(-2,0),
(-1,0),
(0,1),
(0,2),
(1,0),
(2,0),
(3,1) \}. \end{align*}
The sets are shown in Figure \ref{Counterexample} together with coverings by three subsets each: 
\begin{align*}
A_1 &= \{ (0,1),
(0,2),
(1,0),
(1,3),
(2,0)\},\\ A_2 &= \{ 
(1,0),
(2,0),
(2,3),
(3,1),
(3,2)\}, \\ A_1 \cap A_2 &= \{ 
(1,0),
(2,0)
\},\\
B_1 &= \{ (-3,1),
(-2,0),
(-1,0),
(0,1),
(0,2) \},\\
B_2 &= \{ 
(0,1),
(0,2), (1,0),
(2,0),
(3,1) \},\\
B_1 \cap B_2 &= \{
(0,1),
(0,2) \}.	
\end{align*}

It is clear that the similarly indexed (and colored) subsets are pairwise isometric, and so pushouts of the local data should be the same in both cases.  However, the two data sets are not isometric and, in fact, have distinct barcodes as also shown in Figure \ref{Counterexample}, in all dimensions.
\end{ExRef}

In order to state our theorem as a partial resolution to this dilemma, we will need to define the following notion of a partial barcode.

\begin{DefRef}{HTLD}
For a chosen number $\ell$, the \textit{$\ell$-prefix} of a persistent module is the restriction of the functor $H_n \circ \phi$ to the interval $[0, \ell]$.  We will denote this functor by $\ell$-$pH_n$.
If we view the barcode as an unordered collection of intervals with specified times of birth and death of each homology generator, in the form of $\{ (b_{\alpha}, d_{\alpha}) \}$, then  the \textit{$\ell$-prefix} of the barcode is the collection of intervals $\{ (b_{\alpha}, \min \{ d_{\alpha}, \ell \}) \}$.  Visually, one simply erases the part of the given barcode that is beyond the threshold $\ell$.  

We will use the term \textit{simplicial tree} for a simplicial complex that is connected and has no cycles.  A covering of a metric space whose the nerve is a simplicial tree and Lebesgue number at least $\ell \ge 0$ will be called a \textit{tree-like decomposition of rank~$\ell$}.  Suppose each of the covering sets with the subspace metric also has a tree-like decomposition of rank $\ell$.  Then we say that the resulting covering by smaller  sets is a \textit{hierarchical tree-like decomposition of rank $\ell$ and depth 2}.  

Inductively, for a natural number $D$ one defines a \textit{hierarchical tree-like decomposition of rank $\ell$ and depth $D$}. We will refer to the sets that appear in such  hierarchical decomposition and are not unions of other sets as \textit{primary sets}. 
\end{DefRef}

\begin{MainThm}
Given a tree-like decomposition $\mathcal{U}$ of rank $\ell$ and depth $D$ of a finite metric space $M$, the $\ell$-prefix of the persistence module of $M$ based on the Vietoris-Rips filtration can be reconstructed from the $\ell$-prefixes of the persistence modules of each primary set $U$ in $\mathcal{U}$, with the subspace metric, and the $\ell$-prefixes of the persistence modules for a family of intersections $U \cap V$ for the subsets $U$ and $V$ in $\mathcal{U}$.
This follows from the basic technical result which is 
a long exact Mayer-Vietoris sequence 
\mathchardef\mhyphen="2D
\[
\begin{split}
\ldots \longrightarrow \bigoplus_{U, V \in \mathcal{U}} {\ell}{\mhyphen}pH_{n} (U \cap V)
\xrightarrow{\, f_n \, } \bigoplus_{U \in \mathcal{U}} {\ell}{\mhyphen}pH_n (U)
\xrightarrow{\, g_n \, } {\ell}{\mhyphen}pH_n (M) \\
\longrightarrow \bigoplus_{U, V \in \mathcal{U}} {\ell}{\mhyphen}pH_{n-1} (U \cap V)
\longrightarrow \ldots
\end{split}
\]
for a covering of $M$ by two subsets $U$ and $V$ with Lebesgue number at least $\ell \ge 0$.
\end{MainThm}

The easiest to construct and probably the most useful hierarchical tree-like decomposition of depth $D$ can be obtained for any subset of the product of $D$ simplicial or $\mathbb{R}$-trees.  It is explained in the following example.  A further, simplest case of this example is for the trees obtained as triangulations of the real line worked out explicitly in section \refS{EX}.

\begin{ExRef}{hFhv}
Let $T$ be a simplicial tree where each edge is given length 1 with the global metric induced as a path metric.  We fix a vertex $v_0$.  Given another vertex $v \in T$, we define the subset
$\Sh (v) = \{ t \in T \vert v \in [v_0, t) \}$.
Let $B(v,r)$ stand for the open metric ball of radius $r$ centered at $v$ and $S(v,r)$ stand for its boundary sphere.
We also define
\[
\Sh (v,l) = \Sh (v) \cap B(v,l)
\]
for a positive number $l$, and
\[
\Sh (v; l_1, l_2) = \Sh (v, l_2) \setminus B (v, l_1) \setminus S(v,l_1),
\]
for $l_2 > l_1 > 0$.

Given a number $r$ greater than 1, consider the collection of open subsets of $T$ consisting of the ball $B(v_0,2r)$ and the differences $\Sh (v; r-1,3r)$ where the vertices $v$ vary over $S(v_0, (2n-1)r)$ for arbitrary natural numbers $n$.  It is easy to see that (1) this collection is a covering of $T$.  Its nerve is itself a (union of) tree(s) where the vertices can be indexed by $v_0$ and the vertices $v \in S(v_0, (2n-1)r)$, the edges are the pairs $(v,v')$ where $v' \in \Sh (v, 2(n+1)r)$, (2) the diameter of each set in the covering is bounded by $6r$, and (3) the covering has a Lebesgue number at least $r$.  Clearly, intersecting any subset $V$ of $T$ with the produced covering gives a covering of $V$ with exactly the same three properties.  This gives a hierarchical tree-like decomposition of $V$ of rank $r$ and depth $1$

For a product $\Pi$ of $D$ trees, one starts by performing this construction in each of the factors, then takes the product of the covering subsets to create a covering of $\Pi$.  This is a hierarchical tree-like decomposition of rank $\ell$ and depth $D$.  Again, intersecting any subset $V$ of the product with the covering subsets is a hierarchical tree-like decomposition of $V$ of rank $\ell$ and depth $D$.

An \textit{$\mathbb{R}$-tree} or \textit{real tree} in geometry is a geodesic metric space where every triangle is a tripod.  Divergence behavior of geodesics in an $\mathbb{R}$-tree allows to generalize the above constructions in simplicial trees verbatim to $\mathbb{R}$-trees and their products. 
\end{ExRef}

It will be clear from the main body of the paper that for any space with a hierarchical tree-like decomposition of rank $\ell$ and finite depth, there is a finite inductive procedure for reconstructing the $\ell$-prefix of the persistence module of $M$ through a finite number of extension problems from the $\ell$-prefixes of the persistence modules of a selection of intersections between the primary sets.  The details and the implementation of this general algorithm will appear in a separate paper.

The proof of the Main Theorem will be presented in terms of a new theory, singular persistent homology, which we introduce in the next section.  It is true that once the statement is made, the proof can be given entirely in terms of simplicial persistent homology. In fact the two theories are isomorphic when applied to a finite metric space as we will show.  However, let us make three points.  First, it is much easier to think in terms of coverings and related persistence diagrams within the new theory as will become apparent.  This allows to quickly discover the correct statements.  Second, the proofs also appear more natural and easy in the singular theory. Finally, when the complexity of related theories becomes greater, with more general sequences of maps than filtrations, such as in zig-zag or multidimensional persistent homology the transparent nature of the singular theory can be of greater benefit.

\textbf{Acknowledgements.}  The author is grateful to the referee for corrections and insightful comments and suggestions.

\SecRef{Singular Persistent Homology}{BGOAG}

In order to prove the Main Theorem we use an auxiliary theory  that is conceptually better suited for decompositions of metric spaces via coverings.
It is defined for metric spaces that are not necessarily finite or locally finite.  

\begin{DefRef}{Def}
The symbol $\Delta^n$ will denote the metric space with $n+1$ points where the distance between each pair of points is 1.
\end{DefRef}

Let $M$ be a metric space and $R$ be a principal ideal domain.

\begin{DefRef}{Def2}
An \textit{$n$-dimensional singular $\varepsilon$-simplex}  $\sigma \colon \Delta^n \to M$ is an arbitrary set function with the diameter of the image of $\sigma$ bounded by the given real number $\varepsilon \ge 0$. 
Let us denote the set of all such simplices by $S^{\varepsilon}_n (M)$ and the $R$-module of finite chains they generate by $C^{\varepsilon}_n (M)$.

It is easy to see that the usual boundary operation restricts to $C^{\varepsilon}_n (M)$, so we get a chain complex $C^{\varepsilon}_{\bullet} (M)$ for each value of $\varepsilon \ge 0$.  This gives the \textit{discrete singular homology} modules $sH^{\varepsilon}_{n} (M)$.  The ranks of these modules are the \textit{singular Betti numbers} $s\beta^{\varepsilon}_{n} (M)$. 

Clearly the maps induced by inclusions of chain complexes $C^{\varepsilon}_{\bullet} (M) \to C^{\varepsilon'}_{\bullet} (M)$ for $\varepsilon \le \varepsilon'$ give homology homomorphisms which assemble into the familiar persistent module structure for the metric space $M$.  
We will refer to the resulting functor from $[0, +\infty)$ to the category of $R$-modules as the \textit{$n$-dimensional singular persistent homology module} of $M$.  It can be again visually represented by a persistence barcode where the number of bars present at each stage $\varepsilon$ equals the corresponding singular Betti number.
\end{DefRef}

\begin{NotRef}{NAYVDmm}
	We will use the notation $psH_n (M)$ for this functor.  Of course, this new functor is no longer based on a geometric filtration of a space $X$.  Instead, it exploits an algebraic filtration of the chain complex $C^{\infty}_{\bullet} (M)$.
\end{NotRef}

\begin{DefRef}{opers}
It is clear how to do algebra with persistence modules as functors or diagrams modeled on the ordered ray $[0, +\infty)$.  
We will spell out just several operations that are needed in this paper.

A map of persistence modules $h \colon F \to G$ is a natural transformation between $F$ and $G$ as functors, so it is clear what the kernel $K$ and cokernel $C$ of $h$ are in the abelian category of the persistence modules over the p.i.d. $R$.  For example, each $R$-homomorphism $h_{\varepsilon} \colon F_{\varepsilon} \to G_{\varepsilon}$ has a kernel $K_{\varepsilon}$.  There are $R$-homomorphisms $K_{\varepsilon} \to K_{\varepsilon'}$ induced from the structure maps of $F$, and this structure altogether gives a persistence module $K$ which is the kernel of $h$.  

There is also the evident direct sum construction generated by taking direct sums of individual modules at each stage ${\varepsilon}$ and inducing the natural maps between these products as colimits in the module category.
\end{DefRef}

\textit{Historical remark 1.} 
The basic idea here is not new; it goes back to Vietoris's contribution to the development of homology theories for non-polyhedral spaces between 1910 and 1928, building on the pioneering work of L.E.J. Brouwer. The history can be found in Chapter 8 and specifically section 8-6 of \cite{jHgY:61}. At the time, for example, \smash{\v{C}ech} used approximations of continua by nerves of coverings, and Borsuk used embeddings of spaces in geometric models that soon became the basic idea behind shape theory. 
The cycles that Brouwer and Vietoris considered were built out of ordered $\varepsilon$-simplices which were precisely the images of injective singular simplices from Definition \refD{Def2}.  Of course all of this was happening before the comprehensive development of the singular theory by Eilenberg \cite{sE:44} in 1944.

\textit{Historical remark 2.}
One may also consider infinite chains of singular simplices.
A useful additional condition would require that chains are \textit{locally finite} in the sense that each metric ball in $X$ intersects at most finitely many simplices in the chain.  This gives the submodule $C^{\varepsilon,\mathit{lf}}_n (M)$. Of course, when $M$ is a finite metric space, there is no distinction between $C^{\varepsilon}_n (M)$ and $C^{\varepsilon,\mathit{lf}}_n (M)$. 
When $X$ is not compact, the colimit of $C^{\varepsilon,\mathit{lf}}_n (M)$ is essentially the module of uniformly bounded locally finite chains that were defined and used in the work on the Novikov conjecture in $K$-theory by Carlsson/Goldfarb \cite{gCbG:04}.

\textit{Historical remark 3.}  
Persistent homology for non-locally finite metric spaces and finite samples from such spaces in the context of topological data analysis has been studied in several recent papers, for example by Chazal/De Silva/Oudot \cite{fCvDSsO:14} and Adamaszek/Adams \cite{mAhA:17}.

\begin{RemRef}{RFCNHCV}
This comment is related to the historical remark 2 above. The version of locally finite homology in \cite{gCbG:04} is in fact based on a variant of the singular simplex. One can start with a standard simplex $D^n$ in $\mathbb{R}^{n+1}$ and build a singular theory based on maps $\sigma \colon D^n \to M$ which are not necessarily continuous but have the diameter of the image $\sigma (D^n)$ bounded from above by some $\varepsilon$.  We don't include details here but it is not hard to show that the resulting theory is equivalent to the discrete singular homology.  In turn, using the standard subdivision tools, it is possible to prove the following. If $M$ is connected then for any dimension $n$ and any value $\varepsilon \ge 0$ we have $sH^{\varepsilon}_{n} (M) = H_{n} (M)$, where $H_{n} (M)$ is the usual $n$-dimensional singular homology of $M$.
\end{RemRef}

We prove in this section a crucial for computation fact about singular persistent homology.

\begin{ThmRef}{GSASPNBG}
	If $M$ is a finite metric space then $psH_{n} (M) \cong pH_{n} (M)$.
\end{ThmRef}

The isomorphism in this theorem is an isomorphism between persistent modules, so it is an isomorphism of functors from $[0,+\infty)$ to the category of finitely generated $R$-modules.

Sample computations quickly make one suspect that $sH^{\varepsilon}_n (M)$ is isomorphic to $H_n (R_{\varepsilon} M)$ for each value of the parameter $\varepsilon \ge 0$.  Such comparisons are usually proved in algebraic topology by using refinements which are unavailable in finite metric spaces.  Nevertheless, the suspicion is correct.  This is seen by viewing the singular theory as homology of certain simplicial sets.

Let us emphasize that the useful direction in this theorem is certainly the interpretation of the singular persistent homology in terms of the simplicial persistent homology which produces fewer generators on the chain level.  
This fact is used to reformulate the theorem we prove in the next section using the singular theory as the Main Theorem, stated entirely in terms of the simplicial theory.

\begin{proof}
First we define two simplicial sets associated to $M$.  

One is the standard nerve of the Vietoris-Rips complex $R_{\varepsilon} M$ which is denoted $N(R_{\varepsilon} M)$.  This is the simplicial set associated to the poset of the simplices in $R_{\varepsilon} M$.  To spell out what that means, 
$N(R_{\varepsilon} M)_k$ is viewed as all simplicial maps $\Hom ([n],R_{\varepsilon} M)$ from the clique $[n]$ with $n+1$ vertices to $R_{\varepsilon} M$.  The structure maps are the simplicial operators $f \colon [m] \to [n]$ which act by pre-composition
$a \to a \circ f$ for an element $a \colon [n] \to R_{\varepsilon} M$ in $N(R_{\varepsilon} M)_k$.  It is known that the simplicial homology of a complex is isomorphic to the homology of the geometric realization of its nerve.  This isomorphism is induced by the simplicial map from $\vert N(R_{\varepsilon} M) \vert$ to $R_{\varepsilon} M$.  Notice right away that this is in fact a natural transformation between persistence modules and so is an isomorphism of persistence homology modules.
 
The second is the simplicial set $S^{\varepsilon} M$ given by $(S^{\varepsilon} M)_k = S^{\varepsilon}_k (M)$ and the usual for the singular simplicial set face and degeneracy formulas.
It is easily seen to be a Kan complex applying the classical proof verbatim.  The homology of the geometric realization of this simplicial set is the same as the usual homology of the simplicial set built through generating simplicial $R$-modules and further converting to a chain complex, see section III.2 of \cite{pGjJ:09}.  This last construction is precisely how the singular homology $sH_n (M)$ was defined. Each $a \in \Hom ([n],R_{\varepsilon} M)$ is uniquely determined by the values on the vertices.  This clearly gives the isomorphism $t_{\varepsilon} \colon N(R_{\varepsilon} M) \to S^{\varepsilon} M$ which induces isomorphisms on homology.  Moreover, the induced maps for all values of ${\varepsilon}$ give a natural equivalence between $[0,+\infty)$-diagrams which is an isomorphism of persistence homology modules.

Our conclusion is that there is an isomorphism between $pH_n (M)$ and $psH_n (M)$ for all $n$.
\end{proof}

\SecRef{Proof of the Main Theorem}{EP}

Our main theorem is a Mayer-Vietoris statement in singular persistent homology.  We start with a couple of lemmas.

\begin{LemRef}{ONTO}  
For any covering $\{ M_{\alpha} \}$ of a metric space $M$ with a Lebesgue number greater than some nonnegative number $\varepsilon$, the homomorphism
\[
g_n \colon \bigoplus_i C^{\varepsilon}_n (M_{\alpha}) \longrightarrow C^{\varepsilon}_n (M)
\]
induced by inclusions $M_{\alpha} \to M$ is surjective for all $n$.
\end{LemRef}

\begin{proof}
The Lebesgue number assumption guarantees that every singular  $\varepsilon$-simplex, of any dimension, lands in some member of the covering. 
\end{proof}

\begin{LemRef}{INJ}  
Suppose the nerve of the covering $\{ M_{\alpha} \}$ is a simplicial tree.  Choose a total ordering of the indices, so for every edge $\{ \alpha, \alpha' \}$ in the edge set $E$ there is a well defined orientation $\alpha \le \alpha'$.  Then for each edge with the orientation $\alpha \le \alpha'$ we have
the inclusion $i_{\alpha} \colon M_{\alpha} \cap M_{\alpha'} \to M_{\alpha}$ and the inclusion $j_{\alpha} \colon M_{\alpha} \cap M_{\alpha'} \to M_{\alpha'}$.
The homomorphism
\[
f_n \colon \bigoplus_{\{\alpha, \alpha'\} \in E} C^{\varepsilon}_n (M_{\alpha} \cap M_{\alpha'})  \longrightarrow \bigoplus_{\alpha} C^{\varepsilon}_n (M_{\alpha}),  
\]
with the components induced by $i_{\alpha}$ and $-j_{\alpha}$, is injective for all $n$.
\end{LemRef}

\begin{proof}
Since all of the double intersections are disjoint from each other, an element of the kernel forces pairwise cancellations in the images of the homomorphisms induced from $-j_{\alpha}$ and $i_{\alpha'}$.  Finiteness of a sum representing the kernel element guarantees that all summands are 0.
\end{proof}

\begin{ThmRef}{MOBESA}
Given a covering $\{ M_{\alpha} \}$ of $M$ with a Lebesgue number greater than~$\varepsilon$ and nerve a simplicial tree,
there is a long exact Mayer-Vietoris sequence 
\[
\begin{split}
\ldots \longrightarrow \bigoplus_{\{\alpha, \alpha'\} \in E} sH^{\varepsilon}_{n} (M_{\alpha} \cap M_{\alpha'})
\xrightarrow{\, f_n \, } \bigoplus_{\alpha} sH^{\varepsilon}_n (M_{\alpha})
\xrightarrow{\, g_n \, } sH^{\varepsilon}_n (M) \\
\longrightarrow \bigoplus_{\{\alpha, \alpha'\} \in E} sH^{\varepsilon}_{n-1} (M_{\alpha} \cap M_{\alpha'})
\longrightarrow \ldots
\end{split}
\]
\end{ThmRef}

\begin{proof}
This follows from the short exact sequence of chain complexes
\[
0 \longrightarrow \bigoplus_{\{\alpha, \alpha'\} \in E} C^{\varepsilon}_{\bullet} (M_{\alpha} \cap M_{\alpha'})
\xrightarrow{\ f \ } \bigoplus_{\alpha} C^{\varepsilon}_{\bullet}  (M_{\alpha})
\xrightarrow{\ g \ } C^{\varepsilon}_{\bullet}  (M)
\longrightarrow 0
\]
The kernel of $g$ is generated by the elements $x \oplus x'$, where $x \in C^{\varepsilon}_{\bullet}  (M_{\alpha})$, $x' \in C^{\varepsilon}_{\bullet}  (M_{\alpha'})$, and $g(x)=g(-x')$.  So both $x$, $x' \in C^{\varepsilon}_{\bullet} (M_{\alpha} \cap M_{\alpha'})$, and $f = \ker (g)$.
Exactness at the other two terms is obtained from the two lemmas.
\end{proof}

\begin{proof}[Proof of the Main Theorem]
Theorem \refT{MOBESA} can be applied inductively to a hierarchical tree-like decomposition of depth $D$ starting with the computation of persistent homology of primary subsets. The assumption guarantees that the primary subsets form families that are tree-like decomposition of rank $\ell$ of their unions.  Moreover, the unions of the families form a hierarchical tree-like decomposition of $M$ of rank $\ell$ and depth $D-1$.  Since ${\varepsilon}$-$psH$ of each union can be computed using Theorem \refT{MOBESA}, we achieve a reduction in depth.  We inductively exploit hierarchical tree-like decompositions of $M$ and apply Theorem \refT{MOBESA} to finish the computation in $D$ steps.
\end{proof}    

\begin{Ex}
Given a metric space $M$ with a $\lambda$-Lipschitz function $\phi \colon M \to \mathbb{R}$, there is a covering indexed by the integers with a prescribed Lebesgue number $\varepsilon$ constructed as follows.
For any value $\ell > 0$ and any real number $a$, consider the covering of $\mathbb{R}$ by the family of intervals $R_{a,\ell,i} = [ a + i \ell, a + (i+1) \ell + \lambda \varepsilon ]$ indexed by the integers $i \in \mathbb{Z}$.  The overlaps between these closed intervals have size $\lambda \varepsilon$.  Then the covering of $M$ by subsets $ \phi^{-1} [ a + i \ell, a + (i+1) \ell + \lambda \varepsilon ]$ has Lebesgue number $\varepsilon$.  

There is a generalization of this construction for a $\lambda$-Lipschitz function $\phi \colon M \to T$ to a simplicial tree or, indeed, an $\mathbb{R}$-tree.  Select a base point $t_0$ in $T$.  Then consider concentric metric annuli $A_{a,\ell,i}$ in $T$ centered at $t_0$ whose points have distance from $t_0$ fall within $R_{a,\ell,i}$ for positive values of $a$ and $i$.  The inverse images along $\phi$ of the connected components of all $A_{a,\ell,i}$ give the desired covering of $M$ with Lebesgue number $\varepsilon$.
\end{Ex}

We restate the theorem in the case that is useful for our example in the next section.  Suppose that the indexing set of the covering are the integers, and 
the intersections $M_{k} \cap M_{l}$ are empty unless $k = l \pm 1$.

\begin{CorRef}{MIO1}
If the covering $\{ M_{k} \}$ has a 
Lebesgue number greater than $\varepsilon$,
there is a long exact Mayer-Vietoris sequence 
\[
\begin{split}
\ldots \longrightarrow \bigoplus_k sH^{\varepsilon}_{n} (M_{k} \cap M_{k+1})
\xrightarrow{\, f_n \, } \bigoplus_k sH^{\varepsilon}_n (M_{k})
\xrightarrow{\, g_n \, } sH^{\varepsilon}_n (M) \\
\longrightarrow \bigoplus_k sH^{\varepsilon}_{n-1} (M_{k} \cap M_{k+1})
\longrightarrow \ldots
\end{split}
\]
\end{CorRef}

The homomorphisms in the sequence have explicit classical descriptions, so the computation of $sH^{\varepsilon}_n (M)$ and specifically the Betti number $sB^{\varepsilon}_n (M)$ reduces to that of $sH^{\varepsilon}_{n} (M_k \cap M_{k+1})$ and $sH^{\varepsilon}_n (M_k)$ for a finite number of values of $k$.

\begin{CorRef}{MIO2}
	Suppose the metric space $M$ is a finite metric space, so we can use the Vietoris-Rips filtration from Definition \refD{RIPS}.
In view of Theorem \refT{GSASPNBG}, all the statements in this section still hold if we simply replace the the singular homology $sH$ with the simplicial homology $H$ applied to each stage in the filtration.
\end{CorRef}

We want to state this more explicitly in the case the coefficient ring $R$ is a field.
What is needed is the analysis of homomorphisms $f_{\ast}$ because by splitting the long exact sequence into short exact sequences centered around $sH^{\varepsilon}_{\ast} (M)$ we get the description
\begin{equation}
	sH^{\varepsilon}_n (M) \cong \coker (f_{n, \varepsilon}) \oplus \ker (f_{n-1, \varepsilon}). \tag{\dag} \label{cokker}
\end{equation}

This leads to an algorithm for reconstructing the $\varepsilon$-prefix ${\varepsilon}$-$psH_n (M)$ of the usual persistence module when $M$ is a finite metric space and the coefficient ring $R$ is a field.  Before we present this algorithm, we want to make an important remark.

\begin{RemRefName}{POI}{Sufficiency of an $\varepsilon$-prefix}
A necessary feature of the theorems above is that the top scale $\varepsilon$ is fixed a priori and is related to the Lebesgue number of the covering.  This allows to reconstruct only the $\varepsilon$-prefix of the persistence module or persistence barcode.  

We want to argue that there are several common situations in practice where this is not a serious disadvantage. Suppose our data comes from a well-sampled manifold or another geometric shape with tame local behavior.  Then there is a value of the parameter that is known to give a guaranteed reconstruction of homology via homology of the Vietoris-Rips complex.  For example, the main theorems in section 3 of Niyogi/Smale/Weinberger \cite{pNsSsW:06} are of this type.  Another more intrinsic situation is the setting of totally bounded metric spaces due to Chazal/De Silva/Oudot \cite{fCvDSsO:14}. The point is that in these situations the significant persistent homological features of the data set are guaranteed to appear early, likely before the filtration stage $\varepsilon$.  
So sufficiency of the $\varepsilon$-prefix of the barcode will increase with better quality and density of sampled data in conjunction with the greater importance of low dimensional computations.

This is illustrated in the example of a coarser and then finer samples from a circle in $\mathbb{R}^2$ shown in Figure \ref{Sparse}.  The homological features such as connectedness and the single prominent generator in 1-dimensional homology are discernible earlier for the second, finer sample.  It should be clear that the better the density of the sampling the earlier the significant homological features get detected in all dimensions.

\begin{figure}[h!]
      \centering
       \includegraphics[width=0.35\linewidth]{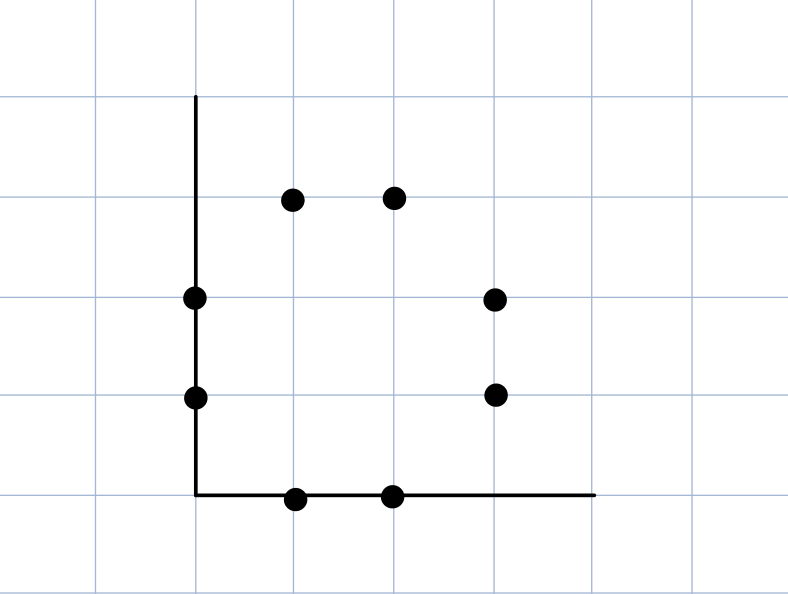} 
       \includegraphics[width=0.85\linewidth]{One.png}
       \includegraphics[width=0.35\linewidth]{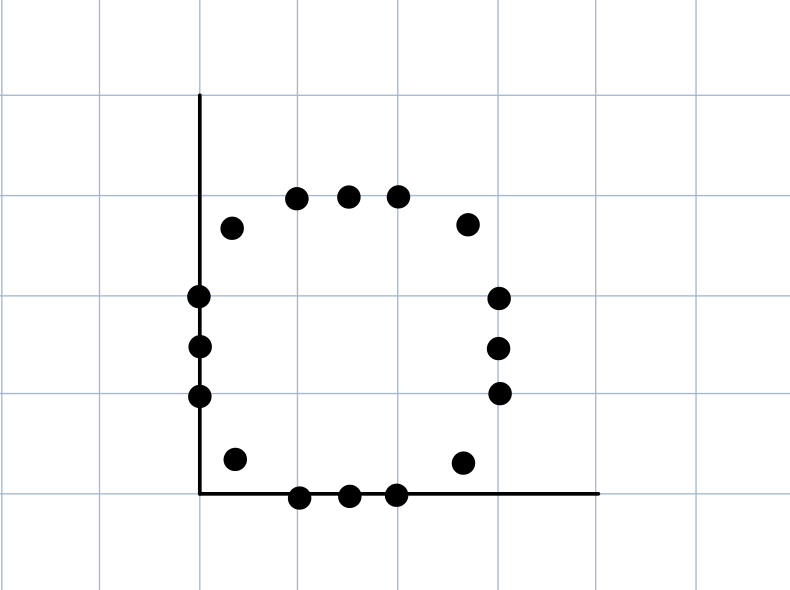}
       \includegraphics[width=0.85\linewidth]{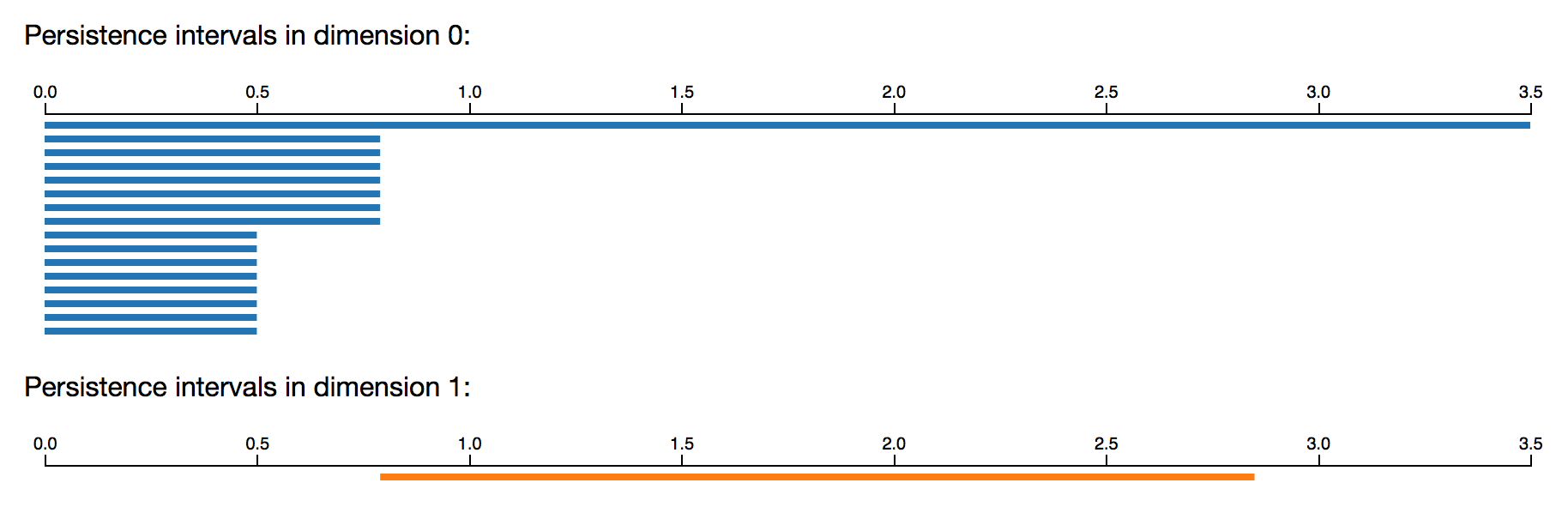}
       \caption{Comparison of coarser and finer samples from a circle.}
      \label{Sparse}
   \end{figure}
\end{RemRefName}

\begin{Alg} \label{1D-alg} 
The following is an explicit algorithm, leveraging Corollaries \refC{MIO1} and \refC{MIO2} and implemented in Figure \ref{fig:1D-code}, for computing the $\varepsilon$-prefix ${\varepsilon}$-$pH_n (M)$ when $M$ is a finite metric space with a 1-Lipschitz function to the real line $\mathbb{R}$.  The value of $\varepsilon$ as well as the dimension $n$ are arbitrary and are fixed from the outset. 

Step 1. Cover the real line by intervals $I_i$ of length $T$ with overlaps of size $\varepsilon < T/2$.  We see that at most two intervals overlap, so the nerve of the covering is 1-dimensional. 

Step 2. The preimages of the intervals $I_i$ along the given 1-Lipschitz function give subspaces $M_i$ of $M$ with a 1-dimensional nerve. 
This gives $M$ a hierarchical tree-like decomposition of rank $\varepsilon$ and depth 1 according to Definition \refD{HTLD}.
In this case the tree is simply the real line.
We denote the lower bound of the index $i$ such that $M_i$ is nonempty by $L$ and the upper bound by $U$. Let us also denote the double intersections by subsets indexed by consecutive integers as $M_{i,i+1}$.

Step 3. Compute the simplicial persistence vector spaces for all pieces ${\varepsilon}$-$pH_n (M_i)$ and ${\varepsilon}$-$pH_n (M_{i,i+1})$ in parallel using one of the available packages such as \texttt{Phat} \cite{BKR:14}, \texttt{Ripser} \cite{uB:18}, \texttt{Eirene} \cite{gH:18}. 

We stress once again that it suffices to compute the usual simplicial persistent homology using the standard packages because of the identification between simplicial and singular persistent homologies in Theorem \refT{GSASPNBG}.

Step 4. Compute $\coker (f_n)$ and $\ker (f_{n-1})$, then the direct sum $\coker (f_{n, \varepsilon}) \oplus \ker (f_{n-1, \varepsilon})$.  This is ${\varepsilon}$-$pH_n (M)$ from the equation (\ref{cokker}). 

Kernels, cokernels, and induced maps between them can be found implemented as methods in all open source software packages listed in step 3.  The same open source libraries can be leveraged to compute ${\varepsilon}$-$pH_n (M)$.

\begin{figure}
\centering
\begin{minipage}[t]{.75\textwidth}
\begin{codebox}
\Procname{$\proc{1D-Mayer-Vietoris}(M,n,\varepsilon)$}
 \li  $\M \gets \proc{1D-Cover}(M,\varepsilon)$
 \li  $\Parfor$ $M({\bullet}) \in \M$
 \li  \Do \proc{$\varepsilon$-pH}$(M({\bullet}))$ \End
 \li  $\Parfor$ $i$ \textbf{from} $L$ \textbf{to} $U$
 \li  \Do \proc{Ind}$(M({i,i+1}) \subset M({i}))$
 \End
 \li  \Do \proc{Ind}$(M({i,i+1}) \subset M({i+1}))$
      \End
        \li  $\Od(n) \gets \proc{Plus}$ $\{ \varepsilon$-pH$(n)(M({i,i+1})) \}$
 \li  $\Os(n) \gets \proc{Plus}$ $\{ \varepsilon$-pH$(n)(M({i})) \}$
 \li  $\Od(n-1) \gets \proc{Plus}$ $\{ \varepsilon$-pH$(n-1)(M({i,i+1})) \}$
 \li  $\Os(n-1) \gets \proc{Plus}$ $\{ \varepsilon$-pH$(n-1)(M({i})) \}$
 \li  $\F(n) \gets \proc{Block}(\Od(n) {:} \Os(n))$
 \li  $\F(n-1) \gets \proc{Block}(\Od(n-1) {:} \Os(n-1))$
 \li  $\mathbf{K} \gets \proc{Ker}(\F(n-1))$
 \li  $\C \gets \proc{Coker}(\F(n))$
 \li  $\mathbf{H} \gets \proc{Plus} (\mathbf{K}, \C)$
\end{codebox}
\end{minipage}
\caption{Pseudocode for computing the relevant segments of the persistent homology Mayer-Vietoris sequence from Corollary \refC{MIO1} and thus its global $n$-dimensional term ${\varepsilon}$-$pH_n (M)$ for a single value of $n$ in parallel. $\proc{1D-Cover}$ generates a cover of $M$ by subspaces $M_i$ and their double intersections $M_{i,i+1}$. The procedure $\proc{$\varepsilon$-pH}$ can be any implementation generating the persistence module for $M_{\bullet}$ for a range of dimensions including $n-1$ and $n$ but which is run only up to the value of the parameter $\varepsilon$.  $\proc{Ind}$ induces persistence diagram maps from inclusions.  $\proc{Block}$ assembles the blocks from lines 5 and 6 into the total matrix of the linear transformation $f_n$ from $\Od(n)$ to $\Os(n)$.  $\proc{Plus}$, $\proc{Ker}$, and $\proc{Coker}$ are the direct sum, kernel, and cokernel procedures described in Definition \refD{opers}. $\proc{Ker}$, $\proc{Coker}$ and matrices for the induced maps are implemented as methods in open source software packages \texttt{Phat}, \texttt{Ripser}, \texttt{Eirene}, and others.
}
\label{fig:1D-code}
\end{figure}
\end{Alg}

The following remarks refer to some notation introduced in Figure \ref{fig:1D-code}.

\begin{RemRef}{NGHCXAQ}
	When computing ${\varepsilon}$-$pH_n (M)$ for a range of dimensions, there is considerable redundancy between computations in adjacent dimensions.  So the array $\F(n)$ would be exactly the same outcome as the array $\F((n+1)-1)$.  Instead of repeating this routine for each dimension, such redundancy can be eliminated by processing the whole range of dimensions at each step and assembling the appropriate outcomes from pairs of consecutive dimensions in the last line. 
\end{RemRef}

\begin{RemRefName}{POIGH0}{Worst case analysis}
This algorithm is most helpful for sparse and uniformly dense data.  These features are common, for example, in graphics data or uniformly sampled data.

To see this point it is instructive to look at the worst case scenario which is all of the data set concentrated in one primary set $M_{i}$.  In this case, there is only one instance of line 3 in Figure \ref{fig:1D-code}.  Then line 5 is the identity map, so eventually 
$\mathbf{H} = \varepsilon$-\textsc{pH}$\,(M({i}))$.
This shows that the algorithm brings no improvement to the computation of $\varepsilon$-{pH}$\,(M)$ in any sense.  At the other extreme are the cases where the range of values in the selected feature of the data is much larger than $R$, and so the algorithm is able to generate many instances of parallel computations in line 3.
\end{RemRefName}

\begin{RemRefName}{POIGH1}{Efficiency}
	The extension steps plus a number of reduced size parallel computations are faster and cheaper in terms of computing resources compared to the head-on computation of the full persistence barcode.  We will analyze a common case in the next section.
	\end{RemRefName}  
	
	\begin{RemRefName}{POIGH2}{Feasibility}
	There exist geometric settings where the Mayer-Vietoris algorithm would not introduce efficiency---for example when the decomposition is modeled on a high-valence tree. Even in cases like that, the ability to parallelize the computation so that each of the barcode computations for covering subsets stays within memory bounds of the processor available for the task makes the whole computation feasible, even if not necessarily efficient. 
\end{RemRefName}

\SecRef{An Example of a Hierarchical Tree-like Decomposition}{EX}

We include an example of a specific practical decomposition which can be applied to subsets of a Euclidean space.  We aim to illustrate both efficiency and feasibility of computing persistent homology of the subset gained from the use of the Mayer-Vietoris algorithm.

It is most natural to divide multi-parameter data according to projections to subsets in one chosen parameter. 
The general kind of parameter for our purposes is tree-based, with partial order, but more commonly the parameter is a coordinate from the totally ordered real line.  The set itself is usually a subset of a Euclidean space where the parameter is one of the Euclidean coordinates. Distributing the computation, one simplifies the computation by applying an algorithm to slices which are reduced in size compared to the total data set. 

In our case this process will be performed inductively using isometric intervals in each real coordinate with overlaps that are at least $\varepsilon$-thick.  
This will guarantee that all slices used in the computation form coverings with a Lebesgue number at least $\varepsilon$.

Before presenting the algorithm, we need to set up a system of notations.

We assume that the given metric space $M$ is a finite set of points embedded as a metric subspace of a Euclidean space of dimension $D$ with the Euclidean metric.  All points are stored as $D$-dimensional vectors.  A simple search can identify the maximal difference $R_j$ between the values of the $j$-th coordinate.  For simplicity we will use $R = \max \{ R_j \}$ but the algorithm can be fruitfully refined by adjusting to the difference in sizes among $R_j$.  Then $M$ is contained in some $D$-dimensional hypercube $[a_1, a_1 + R] \times \ldots \times [a_D, a_D + R]$.  

Suppose we are given a constraint which restricts to $\le p$ the number of computations we are able to run in parallel.
The nature of this constraint can be the  number of processors, or threads, or nodes available at the same time.
Let $\smash{k = \left \lfloor{p^{1/D}}\right \rfloor}$, the largest integer smaller than $p^{1/D}$.  Since $p \ge 1$, this number is positive.

Of course, a larger number $p$ will allow for a more effective parallelization scheme.  The worst case analysis in Remark \refR{POIGH0} corresponds to a single processor being used, with $p=1$.

Now $k$, the floor of $p^{1/D}$, is the number of intervals we use in each Euclidean coordinate.  
Here is a prescription of the intervals in the $i$-th coordinate which achieves all lengths to be 
$T = R/k + \varepsilon$ and the only double overlaps between consecutive intervals of width $\varepsilon$: 
\begin{align}
A_{i,1} &= [a_i, a_i + R/k + \varepsilon], \notag \\ \notag
A_{i,2} &= [a_i + R/k, a_i + 2R/k + \varepsilon], \\ \notag 
\ldots \\ \notag
A_{i,k} &= [a_i + (k-1)R/k, a_i + R + \varepsilon]. \notag
\end{align}
If $\pi_i$ is the projection onto the $i$-th coordinate, it is a 1-Lipschitz function and so produces the strips $S_{i,j} = \pi_i^{-1} A_{i,j}$ which form a covering of $M$ with a Lebesgue number at most $\varepsilon$.

Given an integral vector $x$ with coordinates from $1$ to $k$, there are hypercubes 
\[
C_x = S_{1,x_1} \cap S_{2,x_2} \cap \ldots \cap S_{D, x_D}.
\]
We will generalize this notation as follows.
Let $x$ be an integral vector with coordinates from $0$ to $k$ and with the property that if $0$ appears as a value then all subsequent coordinates must be $0$.  We will use $d(x)$ to denote the highest index for which the value is nonzero.  

\begin{NotRef}{MWSN}
$X(s) = \{ x \in \mathbb{Z}_{\ge 0}^D \mid D-d(x)=s \}$.
\end{NotRef}

Now there is an extension to the previous geometric constructions.  If the value of $j$ is $0$ in $S_{i,j}$, we interpret that set as the union 
\[
S_{i,0} = S_{i,1} \cup \ldots \cup S_{i,k}.
\]
Informally, index $0$ indicates no constraint on the value of the $i$-th coordinate.  So $C_x$ has an interpretation as 
\[
C_x = S_{1,x_1} \cap \ldots \cap S_{d, x_d}
\]
where $d = d(x)$.  There are also sets of the form
\[
C'_x = S_{1,x_1} \cap \ldots \cap S_{d, x_d} \cap S_{d, x_d +1}
\]
when $x_d < k$. 

\begin{NotRef}{UFCC}
	$M_x = M \cap C_x$ and $M'_x = M \cap C'_x$.
\end{NotRef}

What we have here is a hierarchical tree-like decomposition 
$\{ M_x \}$ of $M$ of rank $\varepsilon$ and depth $D$.  The intersections $\{ M'_x \}$ correspond to edges that appear in this scheme. As part of the decomposition, this structure gives a system of 1-dimensional coverings of each $M_x$ for $x \in X(s)$, $1 \le s \le D$. 

\begin{NotRef}{MWSN2}
For $1 \le s \le D$ and any $x \in X(s)$ there is a covering 
\[
\{ M_x \cap S_{D-s+1,i} \mid 1 \le i \le k \} 
\]
of $M_x$ which we denote by $\M (x)$. Another useful way to describe this covering is as $\{ M_y \}$, where 
	\[
	y = (x_1, \ldots, x_{D-s}, i, 0, \ldots, 0) 
	\]
	for $1 \le i \le k$.  When $y$ is as above, we will use the notation
	\[
	y + 1 = (x_1, \ldots, x_{D-s}, i + 1, 0, \ldots, 0) 
	\]
\end{NotRef}   

\begin{ExRef}{JHIQWE}
	The coverings are illustrated in Figure \ref{Decomp}.  Altogether this is an example of a hierarchical tree-like decomposition of depth $D=2$, rank $\varepsilon$ equal to the width of double intersections between cubical sets, and the choice of $k=3$.  Corresponding to $s=1$, we have three possible values of $x \in X(1)$ listed across the bottom of the picture.  The sets $C_x$ for these $x$ are the blue shaded rectangles.  For a data set $M$ contained entirely inside the total region, we have the corresponding subsets $M_x = C_x \cap M$ which form a covering of $M$.  This is a 1-dimensional covering with Lebesgue number $\varepsilon$.  Now let's choose one of these values, say $x = (2,0)$. Corresponding to $s=2$, there are three possible values of $y$ which restrict to $x$ on the first $d-s$ coordinates, so $\M (2,0)$ is the covering of $M_{(2,0)}$ consisting of sets $M_{(2,1)}$, $M_{(2,2)}$, $M_{(2,3)}$ as listed in Notation \refN{MWSN2}.  These are the primary cubical sets colored in red.  Again, this is a 1-dimensional covering with the 
Lebesgue number $\varepsilon$.

\begin{figure}[h!]
      \centering
       \includegraphics[width=0.75\linewidth]{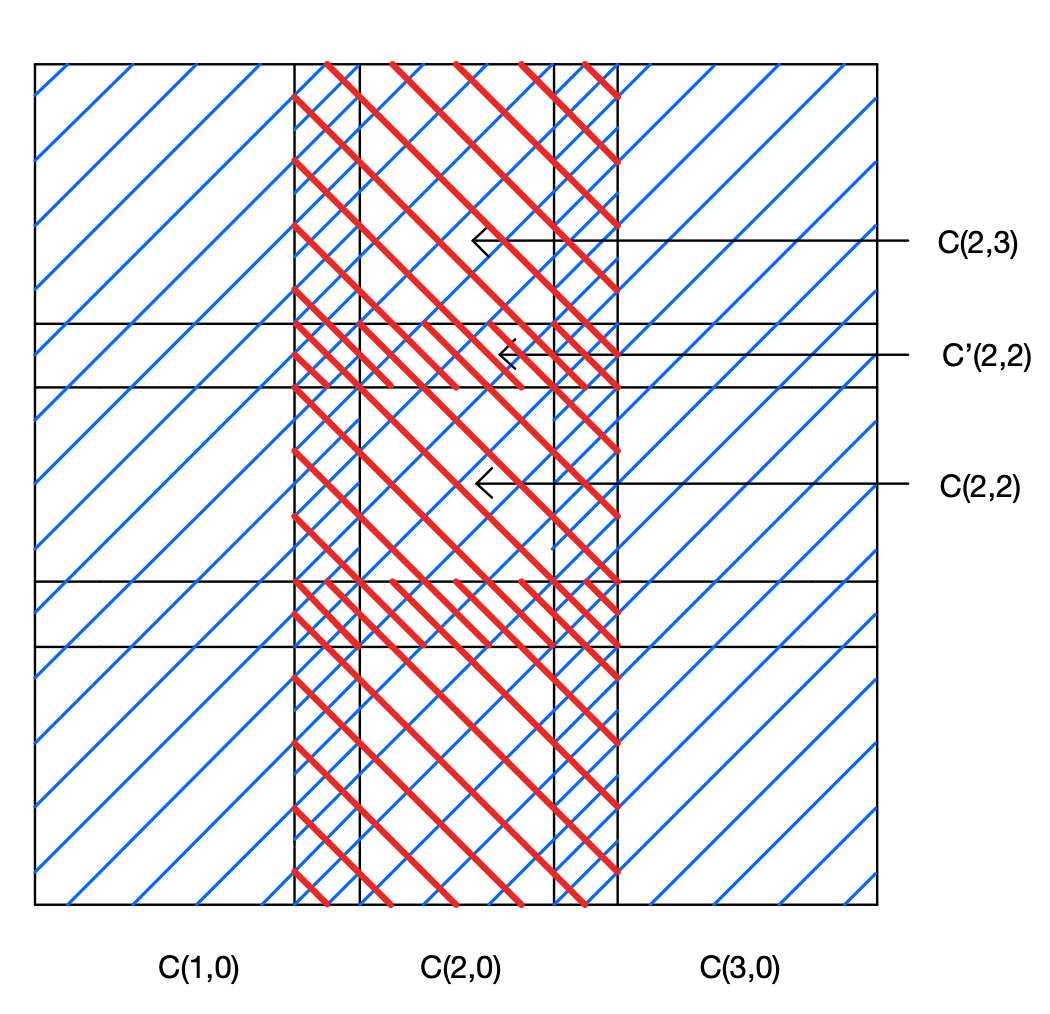} 
       \caption{A hierarchical tree-like decomposition of depth 2.}
      \label{Decomp}
   \end{figure}
\end{ExRef}

\begin{Alg} \label{dD-alg} 
We start with a verbose description, then present a pseudocode.

Step 1. Compute the simplicial persistence vector spaces for all pieces $\varepsilon$-$pH_n (M_x)$, $\varepsilon$-$pH_n (M'_x)$, $\varepsilon$-$pH_{n-1} (M_x)$, and $\varepsilon$-$pH_{n-1} (M'_x)$ for $x$ involving no zeros. The computations can be done in parallel and distributed among available CPUs. This is possible because there are at most $p$ different options for $x$, so $p$ is a sufficient number of processors to perform this step for $M_x$ and then for $M'_x$.

Step 2. Consider all options for vectors $x$ where the $D$-th coordinate is $0$.  Each $M_x$ and $M'_x$ is a union of appropriate $k$ subsets from Step 1 and their $k-1$ nonempty intersections.  Clearly this covering of $M_x$ and $M'_x$ inherits the Lebesgue number bounded by $\varepsilon$.
Applying essentially the procedure $\proc{1D-Mayer-Vietoris}$ we can build $\varepsilon$-$pH_n (M_x)$ and $\varepsilon$-$pH_n (M'_x)$.

Step $S$ (for $3 \le S \le k$).  We proceed inductively applying Step 2 verbatim except restricting to vectors $x$ with $S-1$ last coordinates $0$.  These $M_x$ and $M'_x$ have 1-dimensional coverings by sets from Step $S-1$ with Lebesgue number $\varepsilon$, so the same procedure can be used to compute $\varepsilon$-$pH_n (M_x)$ and $\varepsilon$-$pH_n (M'_x)$.  In the last step, that means computing $\varepsilon$-$pH_n (M_x)$ for the $0$-vector $x$, that is $\varepsilon$-$pH_n (M)$.

\begin{figure}
\centering
\begin{minipage}[t]{.75\textwidth}
\begin{codebox}
\Procname{$\proc{D-Depth-Hierarchical-Assembly}(M,n,\varepsilon,D,p)$}  
 \li  \textbf{build} $X(\mathtt{0})$ as in \refN{MWSN} \End
 \li   $\Parfor$ $x \in X(\mathtt{0})$
 \li  \Do \proc{$\varepsilon$-pH}$(M(x))$ \End
 \li  $\Parfor$ $x \in X(\mathtt{0})$
 \li  \Do \proc{$\varepsilon$-pH}$(M'(x))$ \End
 \li  \textbf{for} \texttt{counter} = $1$ to $D$ 
 \li  \Do \textbf{build} $X(\mathtt{counter})$ as in \refN{MWSN} 
 \li  \Do $\Parfor$ $x \in X(\mathtt{\mathtt{counter}})$
 \li  \textbf{build} $\M (x)$ as in \refN{MWSN2} 
 \li  $\M \gets \M (x)$
 \li  \textbf{evaluate} \proc{$\varepsilon$-pH}($n$)-\proc{Ind}$(M'(y) \subset M(y))$ 
 \li  \textbf{evaluate} \proc{$\varepsilon$-pH}($n$)-\proc{Ind}$(M'(y) \subset M(y+1))$ 
 \li  \textbf{for} \texttt{dim} = $0$ to $n$
 \li  \quad $\Od(n) \gets \proc{Plus}$ $\{ \varepsilon$-pH$(\texttt{dim})(M'(y)) \}$
 \li  \quad $\Os(n) \gets \proc{Plus}$ $\{ \varepsilon$-pH$(\texttt{dim})(M(y)) \}$
 \li  \quad $\F(n) \gets \proc{Block}(\Od(\texttt{dim}) {:} \Os(\texttt{dim}))$
 \li  \quad $\mathbf{K} \gets \proc{Ker}(\F(\texttt{dim}-1))$ if $\texttt{dim} \ge 1$
  \li  \quad $\mathbf{K} \gets 0$ if $\texttt{dim} = 0$
 \li  \quad $\C \gets \proc{Coker}(\F(\texttt{dim}))$
 \li  \quad $\mathbf{H} \gets \proc{Plus} (\mathbf{K}, \C)$
 \li  \quad \textbf{save} $\varepsilon$-pH$(\texttt{dim})(M(x)) \gets \mathbf{H}$ \End
 \li \textbf{save} $\varepsilon$-pH$(n)(M) \gets \varepsilon$-pH$(n)(M(x))$
 \end{codebox}
\end{minipage}
\caption{Pseudocode for a parallel computation of the $n$-dimen\-sional persistent homology term ${\varepsilon}$-$pH_n (M)$ for a finite data set $M$ in $\mathbb{R}^D$ and a single value of $n$ using $p$ processors. 
The time complexity is that of the procedures in lines 3 and 5 which is known to be cubic in the number of simplices in the Vietoris-Rips complexes that get generated by primary subsets. The complexity of all subsequent processes is sub-cubic. The primary sets are the multi-cubes of volume $(R/k)^D$ for $\smash{k = \left \lfloor{p^{1/D}}\right \rfloor}$.  In other words, the reduction of complexity compared to direct persistence computation is proportional to the number $p$ of available processors.
}
\label{fig:dD-code}
\end{figure}
\end{Alg}

The issue of time complexity is different from the issue that is often more important in persistence computations.  Data sets of reasonable size make programs like MATLAB run out of resources even on high-end computers.  Parallelizing the persistence algorithm and distributing the computation among a network of computers with average parameters make possible practical computations that are not even feasible directly, cf. Remark \refR{POIGH2}. 

\SecRef{Discussion}{D}

The major point in this paper is that it is possible to assemble parallel homology computations into partial but often crucial information about the persistence module of a metric space $M$.  The quality of the answer is a decision to be made in the beginning of the computation, but the higher desired quality correlates with greater demands on the number and/or capacity of parallel processors.

The stress here is meant to be on the term ``homology''.  This is distinct from virtually every practical attempt in the literature to parallelize the computation of persistent homology.

There are surely other attempts to parallelize the computation of persistent homology.  They range from general descriptions of strategy to methods that require special arrangements different from our Lebesgue number conditions.
We will briefly survey them and compare.

Bauer, Kerber, and Reininghaus \cite{BKR:14,BKR:15}
are concerned with distributing the matrix algebra involved in the homology computations and optimizations specific to persistence and so are very much transverse to the geometric decompositions in this paper.  These methods are implemented and are part of the toolbox in modern software packages such as \texttt{Phat} \cite{BKR:14}, \texttt{Ripser} \cite{uB:18}, \texttt{Eirene} \cite{gH:18}.
 
Di Fabio and Landi \cite{bDcL:11} work on the level of Betti numbers and detect errors in a Mayer-Vietoris formula for ranks of homology.  They observe that  these estimates can be used in partial matching problems.

Lewis and Zomorodian \cite{rLaZ:14} 
and Lewis and Morozov \cite{rLdM:15} use a very different, hierarchical approach to geometric decomposition of the metric space and use the blow-up complex construction for assembling the data. They build on the work of Carlsson and Zomorodian \cite{aZgC:08} which introduced the blow-up complex for an effective decomposition of the metric space. This is probably the closest in spirit to what is in this paper.  Yet, just as in   Bauer/Kerber/Reininghaus \cite{BKR:14,BKR:15}, the decompositions are used for parallelizing matrix computations of homology rather than the homology computations themselves as we do in this paper.  Another difference is a set of geometric constraints on the coverings that are left by these authors for the user to construct. The problem of finding balanced minimal blowups is NP-hard, and in all cases the resulting multicore algorithm uses a lot of preprocessing.  In contrast, our scheme can be rather canonical and depends only on a specific pattern of overlaps.  We illustrated the scheme with a covering that is always available for our purposes in subsets of a Euclidean space and useful in large well-sampled data sets.

Lipsky, Skraba, and Vejdemo-Johansson \cite{dLpSmV:11} is a very general discussion of how a spectral sequence computation would proceed in favorable situations, organized according to increasing dimension of the Vietoris-Rips complexes.  The algorithmic issues are not addressed.

There is a paper of Zomorodian \cite{aZ:10} that constructs a simplicial set called the ``tidy set''.  This set and its simplifications are used as preprocessing steps for homology computations.  There is no direct relation of that simplicial set to singular homology, and it's unlikely the reductions described by Zomorodian would be useful for singular simplicial sets.

\end{document}